\documentclass[aps,prl,a4paper,showpacs,twocolumn,floatfix,groupedaddress,10pt]{revtex4}
\usepackage{amsfonts}
\usepackage{amssymb}
\usepackage{amsmath}
\usepackage{epsfig}
\usepackage{subfigure}

\usepackage{graphicx}
\usepackage{enumerate}
\usepackage{amsmath}

\newcommand{\ket}[1]{\left|#1\right\rangle}

\newcommand{\ri}[0]{\rightarrow}
\newcommand{\lef}[0]{\leftarrow}
\newcommand{\dow}[0]{\downarrow}
\newcommand{\up}[0]{\uparrow}
\newcommand{\fra}[2]{\displaystyle\frac{#1}{#2}}

\begin{document}

\title{Practical private database queries based on a quantum key distribution protocol}
\date{\today}
\author{Markus Jakobi$^{1,2}$, Christoph Simon$^{1,3}$, Nicolas Gisin$^1$,
Jean-Daniel Bancal$^1$, Cyril Branciard$^1$, Nino
Walenta$^1$, and Hugo Zbinden$^1$} \affiliation{$^1$ Group
of Applied Physics, University of Geneva, CH-1211 Geneva 4,
Switzerland\\$^2$ Humboldt-Universit\"{a}t zu Berlin,
D-10117 Berlin, Germany\\$^3$ Institute for Quantum
Information Science and Department of Physics and
Astronomy, University of Calgary, Calgary T2N 1N4, Alberta,
Canada}

\begin{abstract}
Private queries allow a user Alice to learn an element of a
database held by a provider Bob without revealing which
element she was interested in, while limiting her
information about the other elements. We propose to
implement private queries based on a quantum key
distribution protocol, with changes only in the classical
post-processing of the key. This approach makes our scheme
both easy to implement and loss-tolerant. While
unconditionally secure private queries are known to be
impossible, we argue that an interesting degree of security
can be achieved, relying on fundamental physical principles
instead of unverifiable security assumptions in order to
protect both user and database. We think that there is
scope for such practical private queries to become another
remarkable application of quantum information in the
footsteps of quantum key distribution.
\end{abstract}

\maketitle

\section{Introduction}

As telecommunication gains steadily in importance,
questions of security and privacy naturally arise. Indeed,
private data is stored on a grand scale and has become a
precious commodity. Unfortunately, as a matter of principle
classical information theory is not able to secure privacy
in telecommunication against an unlimited adversary. It was
hence found all the more extraordinary that quantum key
distribution (QKD) allows such ``unconditionally'' private
communication, provided that the two parties trust each
other. However, the more general case of communication
between distrustful parties, who not only wish to protect
their {\itshape common} privacy against eavesdropping, but
also their {\itshape individual} privacy against each
other, is maybe of even greater interest.

Private queries are an important problem of this type.
Imagine that a user, Alice, wants to know an element of a
database held by a database provider, Bob, but does not
want him to know which element she is interested in. Bob in
turn wants to limit the amount of information that she can
gain about the database. In particular, he does not want to
just hand over the whole database, which would trivially
allow Alice to learn her bit of interest without giving any
information on her choice away. It is not hard to imagine
scenarios (e.g. in the financial world) where the
capability of implementing such private queries would be
useful. The information stored in the database may be both
valuable and sensitive, such that Bob would like to sell it
piece by piece, whereas the mere fact of being interested
in an element of the database might already reveal
something important about Alice (e.g. that she is thinking
about buying a certain company). Of course if there was a
cheap way of realizing the task, it would also be
interesting for protecting privacy in online bargaining and
web search, for example, as well as to construct other
interesting cryptographic primitives from it \cite{OTCryp}.

The described task is also known as symmetrically private
information retrieval and as $1$ out of $N$ oblivious
transfer \cite{Rabin}. It has attracted much attention both
in computer science \cite{Computational,Chor} and in
quantum information. Classically, the problem seems like a
logical contradiction. How could a database provider answer
a question, which he is not supposed to know, without
giving any additional information? One might hope that
quantum mechanics could solve this dilemma. Several quantum
protocols were proposed, see for example Refs.
\cite{PractQOT,unbreakableBC}, none of which were found to
offer complete protection for both sides. Indeed, it was
subsequently proven in Ref. \cite{Lo:proof} that the
described task can not be implemented ideally, not even
using quantum physics. The essential assumption in the
impossibility proof is that the protocol is {\it perfectly
concealing}, i.e. that Bob has no information whatsoever
about which database element Alice has retrieved. Rephrased
at the quantum level this is understood as the condition
that the density matrix of Bob's subsystem must be
completely independent of Alice's choice. Ref.
\cite{Lo:proof} shows that under this condition Alice can
always implement an attack based on the Schmidt
decomposition which allows her to read the entire database.
This argument is closely linked to the well-known
impossibility proofs for quantum bit commitment
\cite{Mayers:proof,LoChau:proof}.

Recently, Giovannetti, Lloyd and Maccone \cite{QPQ} pointed
out that very interesting degrees of privacy are achievable
for protocols that are not perfectly concealing, because of
the possibility to catch dishonest parties due to the
errors they introduce, see also
\cite{remark:1,susceptOT,QPQ:exp}. In the protocol of Ref.
\cite{QPQ} Alice encodes her question in a quantum state,
which she sends to Bob. She also sends a decoy state, which
gives her a chance to detect if Bob is cheating. The
security relies on the impossibility to perfectly
discriminate the non-orthogonal question and decoy states,
and on the changes Bob's measurement will introduce as a
consequence. Unfortunately the protocol is very vulnerable
in realistic situations where there are significant
transmission losses, such that Alice has to send the same
question multiple times. If some of the losses are in fact
due to Bob tapping the line, then he can learn Alice's
question without being detected.

\section{Claim}

In this paper we present a new approach to the private
query problem. Our protocol is explicitly not perfectly
concealing in the above sense, so that the impossibility
proof of Ref. \cite{Lo:proof} does not apply. We show that
the following statements hold for our protocol.

\begin{center}
\begin{enumerate}[(1)]

\item {\itshape Database security} is very good. Even for relevant multi-qubit joint measurements Alice's accessible information is restricted to a well-defined small percentage of the database elements. The concrete limits for different attacks are shown in the security discussion. Moreover the additional elements Alice learns are randomly distributed over the database and therefore of little use to
her. In general, database security is ensured by the
impossibility of perfectly distinguishing non-orthogonal
quantum states.

\item {\itshape User privacy} is also very high.
We study several natural attacks and derive a simple limit
on the information Bob can obtain. In general, we show that
the no-signaling principle implies that every malicious
action of Bob will introduce errors and can hence be
detected by Alice - systematic cheating is impossible.
\end{enumerate}
\end{center}

The protocol relies on QKD with changes only in the
post-processing and can hence profit from many of the
advantages of this well understood and commercially
available technology. In comparison to Ref. \cite{QPQ} it
offers the advantage of practical feasibility, in
particular loss-tolerance and scalability to large
databases.

Note that the incorporation of security assumptions such as
the bounded storage model \cite{Storage1} could make the
protocol completely secure, under the condition that those
assumptions are fulfilled. However, even in the absence of
such assumptions, our protocol's basic security is
guaranteed by fundamental physical principles, namely the
impossibility of perfectly discriminating non-orthogonal
quantum states and the impossibility of superluminal
communication.

It should be underlined that we do not propose an ideal
cryptographic primitive, which would furthermore allow one
to construct other ideal cryptographic primitives such as
user identification, bit commitment and coin flipping
\cite{OTCryp}, but a new practical and potentially very
useful application of quantum communication.

Our protocol is similar to the proposal of Bennett,
Brassard, Cr\'{e}peau and Skubiszewska \cite{PractQOT},
which can be interpreted to rely on BB84 QKD \cite{BB:84}.
It is well known that the proposal of Ref. \cite{PractQOT}
is susceptible to a quantum memory attack by the user,
which corrupts database security entirely. The crucial
point is that \cite{PractQOT} is perfectly concealing,
hence Lo's impossibility proof \cite{Lo:proof} implies that
the user can learn the entire database - in this case with
the help of a quantum memory. We show that this type of
attack can be forestalled by using the SARG04 QKD scheme
\cite{SARG} instead of BB84. Then user privacy is slightly
weakened, but the quantum memory attack is no longer
feasible. Moreover the errors a cheating provider
introduces largely guarantee user privacy.

\section{Approach}

In order to better understand our approach it is very
useful to compare it to QKD. In general QKD consists of a
first phase, where a large number of quantum states are
prepared, exchanged and measured, and then a second phase,
where Alice and Bob extract a key from the quantum
communication part with the help of an a priori chosen
coding and interpretation process. The key is then known to
both Alice and Bob entirely and can be used to encrypt the
actual message, which is sent via a classical channel. The
quantum states and the post-processing procedure are chosen
such that the key can not be eavesdropped on without
introducing errors, thus protecting Alice's and Bob's
common privacy.

The basic idea of our protocol is to use QKD in combination
with adequate post-processing to generate an $N$-bit string
$K^f$ that will serve as an {\it oblivious key} \cite{OK}
for a database of $N$ bits. For this purpose, $K^f$ must be
distributed in such a way that (1) Bob knows the key
entirely, (2) Alice knows only a few bits of $K^f$ -
ideally exactly one (database security), and (3) Bob does
not know which bits are known to Alice (user privacy). In
order to use $K^f$ to encrypt the database, Bob adds key
and database bit-wise with a relative shift chosen by Alice
and sends her the encrypted database. The relative shift is
needed in order to ensure that Alice's bit of interest is
encoded with an element of $K^f$ she knows, so that she can
decipher the bit and thus receive the answer to her private
query.

Within our approach, the case of Alice knowing exactly one
bit cannot be realized deterministically. So in general
Alice will know a few bits of $K^f$, which means that
database privacy is good but not perfect. As the number of
Alice's elements is Poisson-distributed, there is also a
small probability of Alice having no bit in the end. The
protocol then needs to be repeated. This can be done
without loss of privacy for either party : The created
string $K^f$ does not contain any information on the
database, so database security is not touched, and likewise
the shift (which maps Alice's known key element onto the
database element she needs) is only communicated once a
correct key has been established. Of course, Alice could
claim to have obtained no element of $K^f$ with the hope of
having more elements after a repetition. However, this
strategy can be made ineffective by choosing the parameters
of the protocol such as to make the case of Alice having no
element very unlikely, cf. also section V.

As already mentioned, the generation of $K^f$ can be based
on QKD techniques. Consider for instance 4-state-BB84-type
QKD. After Bob has sent the states (without further
information), Alice, choosing measurement bases at random,
will measure half of the bits she receives in the correct
basis - without yet knowing for which ones her choice was
correct. When Bob subsequently announces the bases, we have
the situation that (I) Bob knows the entire ``raw key'',
(II) Alice knows half of the bits and (III) Bob can not
know which ones Alice has measured correctly. Alice's
limited information on the raw key can now be further
diluted by adequate processing in order to generate the
oblivious key $K^f$, and this is indeed the way Ref.
\cite{PractQOT} essentially works. However, if Alice has a
quantum memory this protocol is no longer secure. She can
then store the received states and postpone all
measurements until after Bob's announcement. By doing so,
she can learn $K^f$ entirely - there is hence actually no
database security at all.

Fortunately this attack can be largely forestalled rather
easily if one uses a SARG-QKD scheme instead of BB84.
SARG04 uses the same states as 4-state-BB84. The main
difference lies in the attribution of bit values to the
quantum states. Whereas in BB84 one state from each of the
two bases codes for 0, the other one for 1, in SARG04 it is
the basis itself that codes for the bit value. I.e., if Bob
sends a state in the ``up-down'' basis $\updownarrow$ this
signifies a 0, and a state from the ``left-right'' basis
$\leftrightarrow$ means 1. During the post-processing Bob
does not announce which basis he has used for each qubit.
Instead Bob announces the state he has sent plus one state
from the other basis (in random order). Alice is thus faced
with a state discrimination problem that can not be solved
perfectly, i.e. unambiguously and deterministically at the
same time. This slight change has profound implications for
SARG04 QKD \cite{SARG:sec}. Here we show that it is also
very useful for implementing private queries. A simple
protocol based on this approach consists of the following
steps.

\section{Protocol}

\begin{figure}
\includegraphics[width=8.5cm,origin=tc]{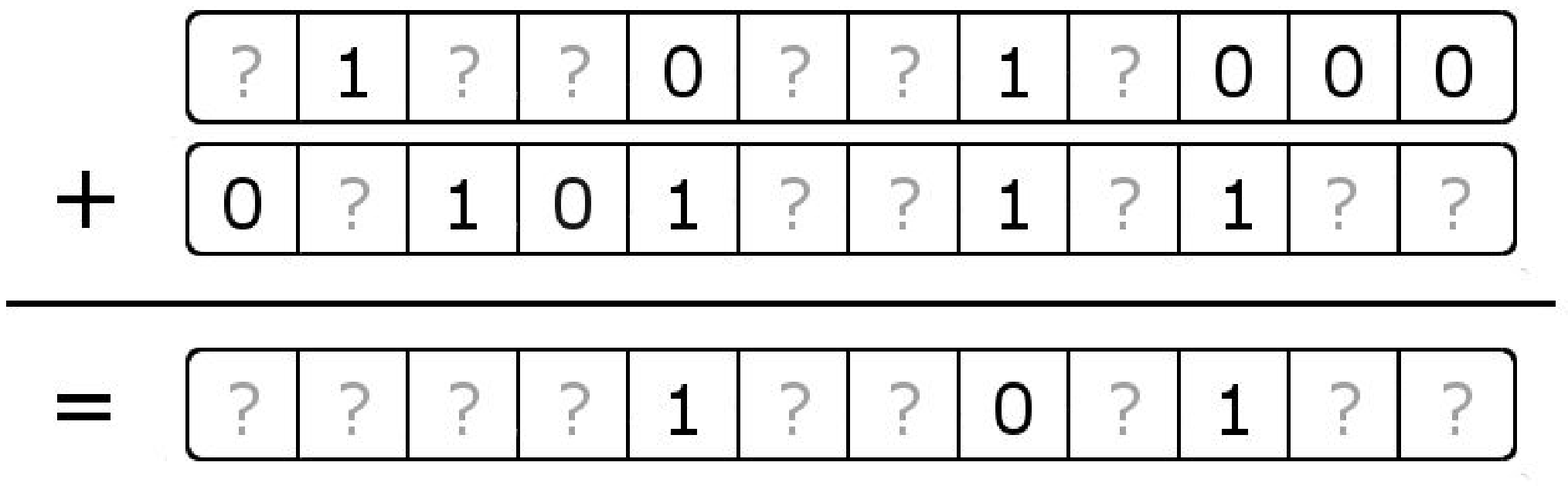}
\caption{\label{Fig:KD} How to reduce Alice's information:
her information on a sum string is lower than that on the
initial strings. Question marks symbolize bits whose value
is unknown to Alice.}
\end{figure}

\begin{center}
\begin{enumerate}[1.]
\item Bob sends a long random sequence of qubits (e.g. photons) in states $\ket{\up}$, $\ket{\ri}$, $\ket{\dow}$ and $\ket{\lef}$. States $\ket{\up}$ and $\ket{\dow}$ code for 0, $\ket{\lef}$ and $\ket{\ri}$ correspond to bit value 1. For instance, to send a bit 1 Bob can prepare a qubit in the state $\ket{\ri}$.

\item Alice measures each state in $\updownarrow$ or $\leftrightarrow$ basis at random. This alone does not allow her to infer the sent bit value.\label{en:measure}

\item Alice announces in which instances she has successfully detected the qubit; lost or not detected photons are disregarded.
The possibility to discard bits does not allow Alice to
cheat, because after step \ref{en:measure} she still has no
information whatsoever on the sent bit values, cf. step \ref{en:interpretation}. As a
consequence, the protocol is completely loss-independent.

\item For each qubit that Alice has successfully measured,
Bob announces a pair of two states: the one that has
actually been sent and one from the other basis, so
$\{\ket{\up},\ket{\ri}\}$, $\{\ket{\ri},\ket{\dow}\}$,
$\{\ket{\dow},\ket{\lef}\}$ or $\{\ket{\lef},\ket{\up}\}$.
If $\ket{\ri}$ has been sent, Bob could announce for
instance $\{\ket{\up},\ket{\ri}\}$. This is exactly as in
the SARG04 QKD protocol \cite{SARG}.
\label{en:lossrestistance}

\item Alice interprets her measurement results of step \ref{en:lossrestistance}. Depending on which basis she has chosen and which result she has obtained she will be able to decipher the sent bit value or not. For instance, if $\ket{\ri}$ has been sent and $\{\ket{\up},\ket{\ri}\}$ was announced, Alice can rule out $\ket{\up}$ only if she has measured in the $\updownarrow$ basis and obtained the result $\ket{\dow}$. She can then conclude that the state was $\ket{\ri}$ and the bit value is 1. Direct measurement as under step~\ref{en:measure} will yield $1/4$ of conclusive results and $3/4$ of inconclusive ones. Both conclusive and inconclusive results are kept. Alice and Bob now share a string which is known entirely to Bob and in a quarter to Alice. \label{en:interpretation}

\item The created string must be of length $k\times N$ (with $k$ a security parameter). It is cut into $k$ substrings of length $N$. These strings are added bitwise in order to reduce Alice information on the key to roughly one bit, cf. Fig. \ref{Fig:KD}.\label{en:end}

\item If Alice is left with no known bit after step \ref{en:end}, the protocol has to be
restarted. The probability for this to occur can be kept
small. See also the discussion in the previous and
following sections.

\item If $K^f$ has been established correctly, Alice will know at least one element of it. Suppose she knows the $j^{th}$ bit $K^f_j$ and wants the $i^{th}$ bit of the database $X_i$. She then announces the number $s=j-i$ in order to allow Bob to encode the database by bitwise adding $K^{f}$, shifted by $s$. So Bob announces $N$ bits $C_n=X_n \oplus K^f_{n+s}$ where Alice can read $C_i=X_i \oplus K^f_{j}$ and thus obtain $X_i$. The shift will hence make sure that Alice's bit of interest is coded with a key element she knows so that the private query can be completed.
\end{enumerate}
\end{center}


\section{Discussion}

Steps 1 to 5 of the above protocol are completely identical
to SARG04 QKD with the only difference that every bit is
kept, regardless if it is conclusive or not for Alice.
SARG04 was initially conceived to make QKD more resistant
to photon number splitting attacks when weak pulses are
used instead of single photons for the sake of practical
feasibility. In our case the use of SARG04 does not only
provide us with the benefits of loss-tolerance,
technological practicability and conceptual closeness to
well-understood QKD, but it also prevents the quantum
memory attack that destroyed the security of the protocol
of Ref. \cite{PractQOT}. Even using a quantum memory Alice
is always confronted with the problem of discriminating two
non-orthogonal quantum states, and will hence always have
incomplete knowledge on the raw key. This lack of
information is subsequently further amplified by step
\ref{en:end}.

Note that following the ''honest'' way of measuring and
interpreting her results Alice will also gain probabilistic
information on non-conclusive bits. If Alice obtains no
result it is with probability $2/3$ because she has chosen
the same basis for measurement as Bob has chosen for state
preparation (which will never yield a conclusive result).
Considering the example of step \ref{en:interpretation},
Alice can obtain the result $\ket{\rightarrow}$ when
measuring in $\leftrightarrow$ both if Bob sent
$\ket{\rightarrow}$ (then with probability $1$) and if Bob
sent $\ket{\up}$ (then with probability $1/2$ only). So,
although $\ket{\rightarrow}$ is not a conclusive result,
Alice can infer that the sent state was $\ket{\rightarrow}$
(bit 1) with probability $2/3$ and $\ket{\up}$ (bit 0) with
probability $1/3$. This additional information can be
diluted to a negligible level by the post-processing of
step \ref{en:end}.

After creation of the raw key of $k\times N$ bits, the
string is divided into $k$ substrings of length $N$.
Following the protocol, after adding the substrings, Alice
will on average know $\bar{n}=N(\frac{1}{4})^k$ bits, where
the number $n$ follows approximately a Poisson
distribution. On the other hand, the probability $P_0$ that
she does not know any bits at all and that the protocol
must be restarted, is
$P_0=\left(1-\left(\frac{1}{4}\right)^k\right)^N \approx
e^{-\bar{n}}$. For large $N$, which is the most interesting
case in practice, it is therefore possible to ensure both
$\bar{n} \ll N$ and small $P_0$ by choosing an appropriate
value of $k$. For instance, for a database of $N=50000$
elements $k=7$ is a choice providing Alice with
$\bar{n}\approx 3$ elements of the final key on average
whereas the probability of failure is only about $5\%$, see
also Tab. \ref{tab:examples}. The case of many repetitions
(which might allow Alice to wait until she obtains a large
value of $n$ by chance) is hence very unlikely. This is
important for the protocol's security. Since the states
sent by Bob do not contain any information about the
database, and since Alice only chooses and communicates the
shift $s$ to Bob once she knows at least one bit of the
final key, a few repetitions will not compromise anybody's
security. Note that even if Alice knows $n>1$ bits of the
oblivious key, she has to pick a single shift $s$, which
means that in general she can only learn one {\it chosen}
element of the database, since the other $n-1$ bits known
to her will be at random positions in the key and thus in
the database.

\begin{table}[tc]
\begin{center}
\caption{Example of possible choices of $k$ for different
database sizes $N$. We show the failure probability $P_0$
and the expected number of elements $\bar{n}$ an honest
Alice will obtain.} \label{tab:examples}
\begin{tabular}{c | c  c  c  c  c  c}

$N$             &   $10^3$      & $5\times 10^3$    &  $10^4$   &  $5\times 10^4$   &       $10^5$  &       $10^6$  \\ \hline
$k$             &           4               &           5           &       6           &       7               &           7           &           9           \\
$P_0$           &           0.020       &           0.008   &       0.087   &       0.047   &       0.002 &         0.022   \\
$\bar{n}$ &         3.91        &           4.88    &       2.44    &       3.05        &           6.10    &           3.81
\end{tabular}
\end{center}
\end{table}

However, the fact that Alice normally obtains additional,
less interesting bits should not be seen only as a drawback
of the protocol, as it also offers an interesting
possibility to enhance her security: Alice can buy the
extra bits in question publicly (as opposed to privately),
in order to compare them with Bob's answers. As explained
in detail in the security section, a cheating Bob will
always lose knowledge on $K^f$. The errors he thus
introduces will then be detectable for Alice. This way what
seems to be a flaw in the protocol can be used to
strengthen user privacy.

\section{Security}

We now turn to the question of which degree of privacy our
protocol offers precisely. We study the most evident
attacks and clarify the way in which two fundamental
physical principles provide the basis for the protocol's
security. While basic attacks are studied and the essential
intuition is given, a complete security analysis remains
work for the future.

\subsection{Database security}

Let us first discuss database security. In general one must
assume that Alice disposes of a quantum memory and is hence
not forced to measure directly as in step \ref{en:measure}.
Instead she can keep the photon and, once Bob has announced
the state pair, apply the optimal {\it unambiguous state
discrimination} measurement \cite{USD:1,USD:2} that will
correctly tell her which of the two announced states has
actually been sent. The success probability of USD is, for
the case of two equally likely states, bounded by
$1-F(\rho_0,\rho_1)$ where $F(\rho_0,\rho_1)$ is the
fidelity between the two quantum states one seeks to
discriminate. Here, Alice's measurement will hence only
work with a success probability of $1-\left|\left\langle
\uparrow \right| \rightarrow
\rangle\right|=1-1/\sqrt{2}\approx 0.29$, only slightly
more than the 0.25 of the direct measurement. In the above
example with $N=50000$ and $k=7$ this will provide her with
$\bar{n}=9.3$ elements on average - only a small gain
compared to $\bar{n}=3$ and very little in relation to
$N=50000$ for such a complex attack. So even using a
quantum memory, individual measurements will not
substantially increase her information on $K^f$. The reason
for this is precisely the fact that our protocol is based
on SARG04 rather than on BB84 coding.

\begin{figure}
\includegraphics[width=8.5cm,origin=tc]{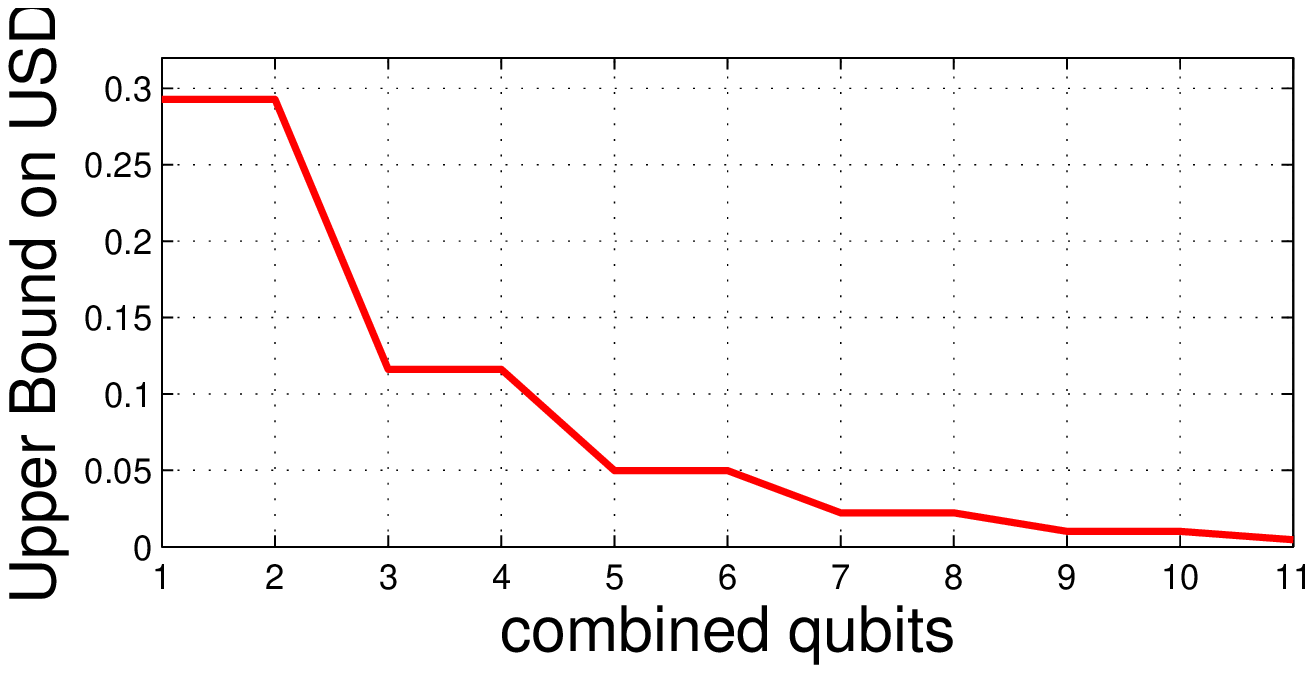}
\caption{\label{Fig:USD} The upper bound on the success
probability of the joint unambiguous state discrimination
(USD) measurement on $k$ qubits declines rapidly with $k$.}
\end{figure}

A more general attack is to store the received photons in a
quantum memory and to postpone all measurements until the
very end of the protocol after step \ref{en:end}, so that
she knows which $k$ qubits contribute to an element of the
final key. The individual bit values of the raw key are
actually of no interest for her. So, instead of performing
the optimal individual measurement on each of the $k$
qubits constituting an element of $K^f$, Alice should
perform a joint measurement. An example for this is
Helstrom's minimal error-probability measurement, i.e. the
measurement that distinguishes two quantum states with the
highest information gain \cite{MaxInfGain:1,MaxInfGain:2}.
In the case of two equally likely quantum states $\rho_0$
and $\rho_1$ the probability to guess the state at hand
correctly is bounded by
$P_{guess}=\fra{1}{2}+\fra{1}{2}D(\rho_0,\rho_1)$, where
$D(\rho_0,\rho_1)$ is the trace distance. For a joint
Helstrom measurement on a bit of $K^f$ one finds this
probability to scale with the number $k$ of added qubits as
$P_{guess}=\fra{1}{2}+\fra{1}{2\sqrt{2^k}}$. So the more
substrings are added to generate the final key, the harder
it is for her to guess the bit value, i.e. the parity of
the $k$ qubits. For example, for $k=7$ Alice will guess a
key element correctly with  $54.4\%$ instead of $50\%$ for
a random guess. Likewise, the success probability of
unambiguously discriminating the two $k$-qubit mixed states
corresponding to odd and even parity declines rapidly with
the number of qubits $k$, see Fig. \ref{Fig:USD}. In
conclusion, it is clear that the {\itshape impossibility to
perfectly distinguish non-orthogonal quantum states} can
effectively protect the database's security and prevent
Alice from knowing a substantial part of it, even when she
uses perfect storage technology and realizes the
theoretically optimal joint measurements. We see that
incorporating a SARG04 state discrimination problem as
vital part of the protocol, the Schmidt attack of Lo's
impossibility proof can be averted. The price to pay is a
protection of the user that is not total. We now turn to
the question of user privacy.

\subsection{User privacy}

As we have discussed above, a not perfectly concealing
protocol, i.e. a protocol where Bob can gain some
information on Alice's choice, is the prerequisite to
prevent her from being able to compromise database security
entirely \cite{Lo:proof}. For the given protocol it may not
be obvious at first sight how Bob can access information on
Alice's choice, in the absence of any classical or quantum
communication from her to him. It turns out that he can
indeed gather information on a bit's conclusiveness, and
hence infer if that particular bit is more or less likely
to be a key element Alice knows.

The simplest attack for Bob is to send other states than he
announces, for instance a state $\ket{\nearrow}$ that is
exactly intermediate between $\ket{\uparrow}$ and
$\ket{\rightarrow}$, while announcing a pair
$\{\ket{\uparrow},\ket{\rightarrow} \}$. Alice's
probabilities to measure $\ket{\downarrow}$ or
$\ket{\leftarrow}$ are largely reduced. Indeed, she will
find a probability of only $14.64\%$ to have such a
conclusive result. Likewise sending the state
$\ket{\swarrow}$ (orthogonal to $\ket{\nearrow}$) while
announcing $\{\ket{\uparrow},\ket{\rightarrow} \}$ will
raise the probability to interpret the result as conclusive
to $85.36\%$. Bob can thus bias the probability of
conclusive results for Alice continuously between the above
limits. However, every such attack will introduce errors,
as Bob cannot predict her outcome with certainty. In the
example above, Alice registering $\ket{\downarrow}$ and
$\ket{\leftarrow}$, i.e. both bit values, are equally
likely events, and Bob's bit error rate will therefore be
as high as $50\%$. This evident example shows that Bob can
gain information on the {\it conclusiveness} of Alice's
bits but will then lose information on the {\it bit values}
she has recorded.

The presented attack is closely related to an attack that
uses entanglement. Bob prepares a state of two qubits
$\fra{1}{\sqrt{2}}\left\{\ket{\up}_A\ket{R_0}_B+\ket{\ri}_A\ket{R_1}_B\right\}$,
where the first qubit is sent to Alice and the second is
kept in Bob's register (with $\langle R_0 \ket{R_1}_B=0$).
Bob announces having sent $\ket{\up}$ or $\ket{\ri}$. Once
Alice has successfully measured and accepted her qubit, Bob
can decide if he wants to measure honestly, i.e. recover
the sent bit value, or gain some information on the
conclusiveness of Alice's measurement. In order to proceed
honestly Bob measures his register in the basis
$\{\ket{R_0},\ket{R_1}\}$, which tells him which of the two
announced states has actually been sent \cite{remark:2}. He
then knows which bit value Alice will record in case of a
conclusive outcome, but has gained no improved estimation
of the likelihood for this to happen. In contrast,
measuring in the
$\{(\ket{R_0}+\ket{R_1})/\sqrt{2},(\ket{R_0}-\ket{R_1})/\sqrt{2}\}$
basis provides him with likelihood information on the
conclusiveness of a bit, but clearly yields no information
at all on the sent bit value.

This second measurement can also be seen from another
angle. If Alice has obtained a conclusive result
(probability $1/4$) Bob's register is in a state
$\rho_c=\begin{pmatrix} 1/2 & 0 \\ 0 & 1/2
\end{pmatrix}$, if Alice measurement was non-conclusive
(probability $3/4$) he has $\rho_n=\begin{pmatrix} 1/2 &
\sqrt{2}/3 \\ \sqrt{2}/3 & 1/2 \end{pmatrix}$. As $\rho_c
\neq \rho_n$ the protocol is not perfectly concealing.
Using the criteria of Refs. \cite{USD:1,USD:2} one can show
that these two density matrices cannot be discriminated
unambiguously for the single-qubit case. The best chance to
guess the state correctly is $85.36\%$, as for the previous
attack. The second given measurement basis does indeed
constitute Helstrom's minimal error probability measurement
\cite{MaxInfGain:1,MaxInfGain:2} for the conclusiveness of
one of Alice's bits. As a matter of fact, one can show
that, given an arbitrary mixed qubit state, the likelihood
to measure a conclusive result will be confined by the very
same bounds ($85.36\%$ and $14.64\%$). No qubit state can
only yield conclusive results upon the above measurement,
or only yield inconclusive results. This individual attack
is therefore optimal, yields information on the bit's
conclusiveness, and completely erases the bit value
information from Bob's register. This last point means that
Bob will not know $K^f$ correctly - a cheating Bob can then
be caught when providing wrong answers \cite{susceptOT}. In
principle these results can be generalized to joint
measurements on several qubits, however, these complicated
attacks are beyond the scope of this paper. Instead we will
now clarify the conceptual reason {\itshape why} it is
impossible for Bob to have both the correct bit value and
conclusiveness information.

Let us suppose that Bob can gain information on the
conclusiveness of one of Alice's elements of the raw key,
either by construction of the sent state, or by some
measurement performed on his register at the end of the
protocol. Let us characterize this information by $p_c$,
the probability with which Bob correctly guesses that Alice
has a conclusive result. (Remember that this likelihood is
physically bounded by $0.1464\leq p_c\leq 0.8536$ if a
single qubit is sent.) Let us also assume that, either by
construction of the state or by some second measurement,
Bob can also guess the bit value $b$ Alice has recorded (if
her measurement was conclusive) and is correct about it
with the probability $p_b$. Recalling the way Alice
interprets her measurement results in step
\ref{en:interpretation} of the protocol, it is clear that,
if Bob correctly guesses that Alice's result was indeed
conclusive and correctly guesses which bit value she has
obtained, then he also correctly guesses which measurement
basis she has used for this qubit in step \ref{en:measure}.
However, since there is no communication whatsoever from
Alice to Bob about her choice of basis, the {\itshape
no-signaling principle} dictates that his probability to
guess her basis correctly has to be equal to 1/2. Otherwise
the procedure would allow Alice to send signals to Bob that
are faster than the speed of light. This immediately
implies the bound
\begin{equation*}
p_c\times p_b \leq 1/2.
\end{equation*}
The inequality arises because even for inconclusive results
Bob has a chance to guess Alice's basis correctly. This
simple upper bound illustrates the crucial point:
 Whenever Bob tries to alter
the conclusiveness probability of certain bits in order to
better judge which bits of $K^f$ are (un)known to Alice, he
will necessarily lose information on the bit value Alice
records, in order to comply with the no-signaling
principle. This introduces errors in $K^f$ and hence also
in the encrypted database, i.e. he will run the risk of
giving wrong answers.

This shows that our protocol is cheat-sensitive in the
spirit of Refs. \cite{QPQ,susceptOT}. In our scenario, Bob
sells his database bit by bit. Systematic cheating and
hence giving wrong answers will ruin his reputation as a
database provider. As we already mentioned above, one can
now even make use of the fact that Alice normally obtains
additional database elements. If she buys those elements
from Bob in a regular, non-private way, she can use them to
check Bob's honesty \cite{remark:3}. By doing so, Alice has
a powerful prompt privacy check at hand. One can thus turn
what seems a flaw into an advantage, in order to make full
use of the privacy, which, as we have seen, is guaranteed
by the {\itshape impossibility of superluminal
communication} in quantum physics.

\section{Outlook \& Conclusions}

The above discussion has shown that practically very
interesting levels of privacy in database queries can be
achieved for both sides. The security of the presented
protocol relies on fundamental physical principles (the
impossibility to deterministically discriminate
non-orthogonal states, and the impossibility of
superluminal communication), rather than on assumptions on
quantum storage limitations \cite{Storage1}, mathematical
complexity \cite{Computational} or non-communication
between severs in multi-server protocols \cite{Chor}.

We have already emphasized that the protocol is completely
loss-resistant. We believe that error correction is
possible as well. This requires additional classical
two-way communication and still needs to be elaborated in
more detail. Moreover, it is clear that the protocol can be
implemented with weak coherent pulses as well. The
acceptable amount of loss then depends on the mean photon
number per pulse, in order to safeguard database security.
High mean photon numbers largely facilitate unambiguous
state discrimination for Alice, if one assumes that she is
in control of the transmission line. Finally, it is
possible to improve database security by more sophisticated
post-processing, e.g. by taking a couple of strings created
in our probabilistic protocol (with $P_0\ll 1$) and
allowing Alice to combine them, i.e. to freely choose
relative shifts to add them bitwise. Simulations show that
she will be left with knowing exactly one bit of the final
key with overwhelming probability. Both error correction
and the described way of achieving tighter database
security complicate the security analysis due to the
necessary two-way communication.

The proposed protocol can be realized with any existing QKD
system that is compatible with the SARG04 protocol. Besides
ensuring loss tolerance, this also makes it easy to scale
up to large databases. We hope that our proposal will
stimulate further work to clarify the open questions.
Besides a more in-depth study of its security, these
include the optimal classical procedures for oblivious key
generation and error correction. We think that there is the
potential for private queries to become a genuine
application of quantum information technology in the
footsteps of QKD.

We thank G. Brassard, V. Giovannetti, S. Hastings-Simon, U.
Herzog, L. Maccone, S. Pironio, C. Schaffner, D. Stucki, S.
Wolf, and J. Wullschleger for useful discussions and
insightful comments. Financial support by the Swiss NCCR-QP
and the European ERC-AD Qore is gratefully acknowledged.
C.S. was supported by an NSERC Discovery Grant.


\begin{thebibliography}{100}
\bibitem{remark:1} This implies that one has to accept a certain trade-off
between user privacy and database security. This has been
studied quantitatively for bit commitment \cite{BCtrade-off}.
\bibitem{remark:2} This procedure is equivalent to rolling a quantum die in order to randomly decide which state to send.
\bibitem{remark:3} Once Bob has sent the encrypted database over a classical channel, he must have measured his quantum register. There is hence no more way to decide to measure certain bits honestly and others not. The cheating is then detectable in the bit error rate of the key, i.e. in the wrong answers he gives.
\end{thebibliography}
\end{document}